\documentstyle[12pt,epsf]{article}

\newcommand \be  {\begin{equation}}
\newcommand \ee  {\end{equation}}
\newcommand \bea {\begin{eqnarray}}
\newcommand \eea {\end{eqnarray}}

\begin{document}

\title{Numerical study of the Ising spin glass in a magnetic field}

\author{Marco PICCO$^{(a)}$ and Felix RITORT$^{(b)}$ \\[1.5em]
{\small $(a)$: LPTHE}\thanks{Laboratoire associ\'e No. 280 au CNRS.}\\
{\small Universit\'e Pierre et Marie Curie}\\
{\small and Universit\'e Denis Diderot}\\
{\small Bte 126, 4 Place Jussieu}\\
{\small 75252 Paris CEDEX 05, France}\\
{\small $(b)$: Dipartimento di Fisica and INFN}\\
{\small Universit\`a di Roma {\it Tor Vergata},}\\
{\small Viale della Ricerca Scientifica, 00173 Roma, Italy}\\[0.5em]
%{\small $(c)$: Departament de Fisica Fonamental,}\\
%{\small Universitat de Barcelona,}\\
%{\small Diagonal 645, 08028 Barcelona, Spain}\\
{\footnotesize \tt picco@lpthe.jussieu.fr/ritort@roma2.infn.it}\\[1.0em]}
\date{March 1994}
\maketitle

\vfill

\begin{flushright}
  {\bf Preprint ROM2F/94/06}\\
  {\bf Preprint PAR-LPTHE 94/14}\\
  {\bf cond-mat/9403077}\\
\end{flushright}

\newpage

\begin{abstract}
We study the order parameter distribution $P(q)$ in the $4d$ Ising spin
glass with $\pm J$ couplings in a magnetic field.  We also compare these
results with simulations for the infinite ranged model (i.e. SK model.)
Then we analyse our numerical results in the framework of the droplet
picture as well as in the mean field approach. 

\end{abstract}

\vfill
\newpage

%_______________________________________________________
This work is devoted to the study of spin glasses in presence of a
magnetic field. During the last ten years, a large number of works has
been devoted to the study of spin glasses \cite{1}. One of the problems which
still remains unsolved is to understand the effect of a magnetic field
on the spin glass phase. The mean-field theory predicts that the
spin glass phase will survive to the application of a magnetic field
below the de Almeida-Thouless critical line (AT line)
\cite{2a}. In the most general case, the main effect should be the
destruction of a large number of equilibrium states with a reshuffling
of the free energies for the remaining ones. To our knowledge, even at the mean
field level, a numerical test of the theoretical
predictions of the replica symmetry breaking solution with magnetic
field has never been done. Such a test would be interesting because 
it would give support to 
the Parisi ansatz as a correct solution to mean-field theory \cite{2}.

For the short-range models case, there is still much controversy.
In fact, there is no precise theoretical prediction.  
A usual $\epsilon$ expansion near dimension $6$ will run in trouble 
because it is not known how to find a non trivial fixed point
\cite{3}. Phenomenological models like those developed by D.~S.~Fisher
and D.~A.~Huse \cite{4} predict that the spin glass phase disappears 
for a finite magnetic field. But this result is 
a consequence depending on some assumption on the real nature of 
the low temperature spin glass phase.
The most recent studies of the AT line were done in Monte Carlo simulations.
In these works, the main points of interest were the curves of constant 
non-linear susceptibility in the $h-T$ plane
\cite{5} or its divergence when approaching the AT line using
finite-size scaling methods \cite{6}. Nevertheless the first approach is
very indirect and the second one could be plagued by strong
corrections to the simple scaling.

In this work we have tried to understand the spin glass with magnetic
field by studying the $P(q)$ order parameter function obtained by means of
the Monte Carlo method using the heat bath algorithm. We will first
present a brief discussion on the general theoretical
predictions of this problem and numerical results for the mean-field
theory. Then we will present the numerical results obtained for the 
$4d$ $\pm J$ Ising spin glass.  It is well established that this
model has a finite $T_c$ and it has been intensively studied \cite{7}.
%(in the 3d case it is possible that there is not a finite temperature
%phase transition \cite{7a}).  

The $d$-dimensional Ising spin glass model of interest, with $\pm J$ 
couplings, is defined by the following hamiltonian
\be
H=-\sum_{(i,j)}\,J_{ij}\sigma_i\sigma_j\,-\,h\sum_i\sigma_i
\label{eq1}
\ee
The couplings $J_{ij}$ are quenched variables with equal distribution values 
$\pm 1$. The interaction is restricted to
nearest neighbors and $h$ is the magnetic field. The Ising spins
$\sigma_i$ take two possible values $\pm 1$ and live in a
$d$-dimensional hypercubic lattice with periodic boundary conditions. It
is very useful to consider discrete couplings $J_{ij}$ in the
hamiltonian because this speeds up the updating of the spins in the Monte
Carlo numerical simulation and their discreteness should not be relevant
for the physics at least for not too low temperatures. In the limit
$d\to\infty$, one expects to converge to mean-field theory, i.e. the SK
model \cite{7b}.  In the SK model, all spins interact among them
and the couplings $J_{ij}$ are normalized by a factor $1/\sqrt{N}$ where
$N$ is the number of spins.

We consider two identical copies of the system eq.(\ref{eq1}), i.e with
the same realization of the bond disorder $J_{ij}$ \cite{8}. Let us call
them $\lbrace \sigma_i\rbrace$ and $\lbrace \tau_i\rbrace$.
The overlap $Q$ among the two copies is defined by
\be
Q=\frac{1}{N}\sum_i\,\sigma_i\tau_i
\label{eq2}
\ee
from which we can construct the order parameter function $P(q)$
\be
P(q)=\overline{\langle\delta (q-Q)\rangle}
\label{eq3}
\ee
where $\langle...\rangle$ and $\overline{(...)}$ mean the usual
statistical Gibbs average over configurations and the average over the
quenched disorder respectively.

In mean-field theory, below the critical temperature and at zero magnetic
field, there exist an infinity of equilibrium states, all of them having a 
different statistical weight. Furthermore, all these states have zero
magnetization and no state is particularly selected if we apply a
magnetic field. In fact, an infinity of equilibrium states still remain
and the spin glass phase survives in a magnetic field. At zero magnetic
field the order parameter distribution $P(q)$ is symmetric under the
exchange $q\to -q$. The magnetic field breaks this symmetry and the
$P(q)$ is expected to be non-zero only if $q>0$. This means that half of
the states have been suppressed by the magnetic field. 

Close to
$T_c$, the free energy of the SK model with magnetic field can be
approximate by
\be
f=\tau \sum_{a<b} Q_{ab}^2- \frac{1}{6}\,Tr Q^3-
\frac{1}{12}\sum_{a<b}\,Q_{(ab)}^4
-h^2\sum_{a<b}Q_{ab}
\label{eq4}
\ee
where $\lbrace Q_{ab};1\le a,b\ge n\rbrace$ is the order parameter and $n$
the number of replicas.  The equilibrium solution of the free energy in the
limit of infinite order of replica symmetry breaking gives a function
$q(x)$ defined in the interval $(0,1)$ \cite{2}. This function $q(x)$ is
the analytical continuation of the matrix $Q_{ab}$ in the limit $n\to
0$. Close to $T_c$ one gets a $q(x)$ with two plateaus in the regions
$0\leq x\leq x_{min}$ and $x_{max}\leq x\leq 1$ ($x_{min}\leq x_{max}$)
with respective values $q_{min}$ and $q_{max}$ ($q_{min}<q_{max}$). Between
$x_{min}$ and $x_{max}$, $q(x)$ increases with $x$.  $q_{max}$ is nearly
independent on the field but $q_{min}$ increases with a power of $h$
smaller than $1$ ($q_{min}\sim h^{\frac{2}{3}}$). Using the static chaos
approach to spin glasses \cite{11} it has been suggested that the effect of
the magnetic field is the progressive suppression of all equilibrium states
$\alpha$ such that their overlap $q_{\alpha \beta} <
q_{min}\,\forall\beta$.  This corresponds to cutting some branches of the
ultrametric tree.  Using the relation $P(q)=\frac{dx(q)}{dq}$ \cite{8a} one
finds that $P(q)$ is given by a continuous part $P_0(q)$ which is non zero
inside the interval $(q_{min},q_{max})$ and two singularities at the
extremes of this interval
\be
P(q)=P_0(q)+a\delta(q-q_{min})+b\delta(q-q_{max})
\label{eq5}
\ee

We
have simulated the SK model at $T=0.5$ and $h=0.3$ (the corresponding
field at the AT line at that temperature is $h\simeq 0.57$). We
simulated three different sizes $N=320,1048$ and $3200$. For these
sizes we were able to reach equilibrium for near all samples after
100000 Monte Carlo steps for the largest size. Then statistics was
collected over several hundred thousands of Monte Carlo steps. The main
source of fluctuations comes from the finite number of samples because
the $P(q)$ is strongly non self-averaging. Self-averageness is restored
when the AT line is reached by increasing the field. The number of samples
is $500,30$ and $20$ respectively. Even though the numbers of samples are
small, they are large enough to show the
qualitative behavior of the $P(q)$. The results are shown in Fig. 1.
The $P(q)$ begins to display two singularities for sizes
of several thousands spins. For smaller sizes, we only found one peak
plus a long tail which extends down to the region of negative overlaps.
According to the Parisi solution to mean-field theory, 
$q_{min}$ should match the correct value at infinite order of replica
symmetry breaking in the infinite-size limit.  We can then compute the
position of both singularities, at least at first order of replica
symmetry breaking and this gives $q_{min}=0.45$, $q_{max}=0.63$.  This is
in agreement with our numerical results.  Other good estimates for 
$q_{min}$ and 
$q_{max}$ are also obtained using the PaT (Parisi-Toulouse) hypothesis
\cite{9} (which is a very good approximation at least close to $T_c$).
This approximation predicts that
$q_{min}(h,T)=q(h,T_{AT}(h))$ and $q_{max}(h,T)=q(h_{AT}(T),T)$ where
$T_{AT}(h)$ and $h_{AT}(T)$ are the equations for the AT line. Computing
these values, one gets $q_{max}\simeq 0.64$ and $q_{min}\simeq
0.437$. This is also in agreement with our simulations.

Now we return to the 4d case. There exists two possible scenarios that we 
want to compare. First, from the droplet models \cite{4} 
it is expected that all
excitations of droplet of sizes larger than a certain length $\xi$ will
be suppressed by the field. The dependence of this correlation length in 
function of the magnetic field is given by
\be
\xi\sim (q_{EA}\,h^2)^{\frac{1}{2\theta-d}}
\label{eq6}
\ee
with $q_{EA}$ the Edwards-Anderson order parameter and $\theta$ the
thermal exponent which gives the characteristic energy scale
$L^{\theta}$ of droplet excitations of typical size $L$. This exponent
$\theta$ should be approximately $\frac{d-3}{2}$ (as emerges from
numerical studies of chaos in spin glasses \cite{10,11}.) For sizes much
larger than $\xi$ it is expected that the $P(q)$ will be strongly peaked
around a unique value of $q$.

%According to mean-field theory we expect that infinite equilibrium
%states should survive. Because all these states have a finite weight
%probability in the infinite size limit we expect that $P(q)$ should
%converge to something resembling eq.(\ref{eq5}). Now we also expect that
%applying a magnetic field the system will resemble itself up to a
%certain correlation length. The difference is that now this correlation
%length does not give the typical size of excitations which have been
%suppressed by the field but the typical distance during which one equilibrium
%state resembles to another. Then typically states with $q$ negative  are
%suppressed with a probability which do not vanishes exponentially with
%the volume $\xi^d$ but much more slowly. 

We have simulated $L=3,5,6,8$ in a 4d lattice with periodic boundary
conditions with $\pm J$ couplings. Simulations were performed at
$T=1.2\,(\sim 0.6\,T_c)$ and $h=0.4$. The number of sample are
$320,128,100,50$ respectively. From finite-size scaling studies
\cite{6}, we expect to be within the spin glass phase if there is an AT
line. It is not very 
difficult to reach the equilibrium in case of $L=3,5$. $100000$ Monte Carlo
steps were enough after a slow cooling procedure.  Now we will try to
convince the reader that we effectively thermalized for $L=6,8$.  To this
end we performed a simulated annealing of half a million of Monte Carlo
steps from the high to the low temperature phase at constant magnetic
field. After that, statistics was collected over the next half a million
Msteps. During this collecting, we computed the four moments of the $P(q)$
distribution which show no apparent drift in time. To increase the
statistics we simulated in parallel eight identical copies of the system
computing the four overlaps among four different pairs at each Monte Carlo
step.  The fact that the magnetic field tries to align the spins helps in
the thermalization procedure. This is the reason why we were able to
thermalize over a scale of time of half a million of Monte Carlo steps
which would be probably insufficient at zero magnetic field.  Figure 2 shows
the numerical results for the $P(q)$. We can immediately notice that there
is no singularity at $q=q_{min}$ if we compare to the previous figure for
the SK model.

Looking at this results it is difficult to draw a definite conclusion
on what is the correct scenario in $4d$ Ising spin glasses. Two facts
are interesting to point out. The first one is the existence of a long
tail for sizes up to $L=6$ which extends down to negative
overlaps. So, $P(q=0)$ is finite which means that reversal of compact
domains of characteristic size $L$ are still present with a finite
probability.  Within the droplet model, we can estimate how domain
excitations of typical size $L$ are suppressed by the magnetic field.
The effect of the magnetic field depends on the regime in which the
system is, either $L >> \xi$ or $L<< \xi$, $\xi$ being given by
eq.(\ref{eq6}). We can estimate the value of $\xi$ by using numerical
simulations of static chaos \cite{11}. A typical value of order $5$ is
obtained. If $L >> \xi$ we expect that the droplets excitations of
size $L$ are suppressed with a factor $\simeq exp(-\beta \chi ({L})^d
h^2)$ respectively to the case $h=0$ ($\chi$ being the linear
susceptibility.) So, in this regime, the tails would be suppressed.
Unfortunately, we are in the regime where $\xi \sim L$. As a lower
bound, when $L<< \xi$, tails are suppressed with a factor $\simeq
exp(-(L)^{d\over 2} h^2)$. In our range of sizes, this factor is of
order $10^{-1}$ which is smaller than what we can see on our plots
($P(q=0)$ being $0.3$ at zero magnetic field, we would expect $P(q=0)
\sim 0.03$ \cite{7}.)  The second fact regards the absence of a second
peak of $P(q)$ at $q_{min}$. Presumably, such a peak could appear for
larger sizes. In the case of SK model, the $q_{min}$ peaks already
arise for size of order $1000$ spins, as oppose to the $4d$ case.  One
possible reason for such a difference reside in the fact that, for the
$4d$ case, we can be very close to the AT line. A second reason is
that we can surely expect stronger finite size effects than in the
mean-field case. For instance, the singularity in $P(q)$ for
$q=q_{max}$ is less pronounced as can be seen in numerical simulation
\cite{7}.

In order to reach more definite conclusion, we need to study larger
sizes lattices. In practice, such a task is very difficult because
for larger sizes lattices we are not able to thermalize. In fact, we
have performed numerical simulations for $L=10$ and $L=12$. Despite
that these are non equilibrium results, interesting hints can be
obtained. In Figure 3 we show the $P(q)$ distribution. Starting from
uncorrelated configurations, the overlap among two copies grows with
time. In several cases it remains stacked in a value of $q$ close to
0.4 giving two singularities for the $P(q)$ distribution. 
This indicates that we are in the good region in the $h-T$ plane 
in order to test if there exists a spin glass phase. 

Still, in order to have a more definite conclusion, we need to
take advantage of new numerical simulation techniques like the simulated
tempering \cite{Pa92}. This method has revealed much effective for the
$2d$ \cite{Ke94} and $3d$ \cite{Ma94} Ising spin glasses. 
A work using such techniques is under progress.

Summarizing, we have studied the 4d Ising spin glass with magnetic
field. For comparison, we have also simulated the SK model. This is
also a test of the Parisi solution to mean-field theory and our
numerical results are in agreement with it. In the $4d$ case we
present results for lattice size up to $L=8$. Then we tried to
interpret them in the mean-field picture and the droplet one. It seems
that the effect of the magnetic field is weaker than what
droplet picture predicts. Non thermalized results for larger sizes
suggest that fully equilibrated simulations should be able to select
in a definite way between these two pictures.  We hope that using
numerical techniques like simulated tempering should be able to decide
the question in the near future.

\section*{Acknowledgments}
We thank D. Huse and G. Parisi for very interesting suggestions. F.~R.
has been supported by the EC fellowship B/SC1*/915198 and the INFN. M.~P.
has been partially supported by the EU Science grant
SC1*0394 and EU Capital Humain et Mobilit\'e grant - Euronetwork ERB
4050PL92 0982/0035.

\vfill\eject 
\newpage
{\bf Figure Captions} 
\begin{itemize} 
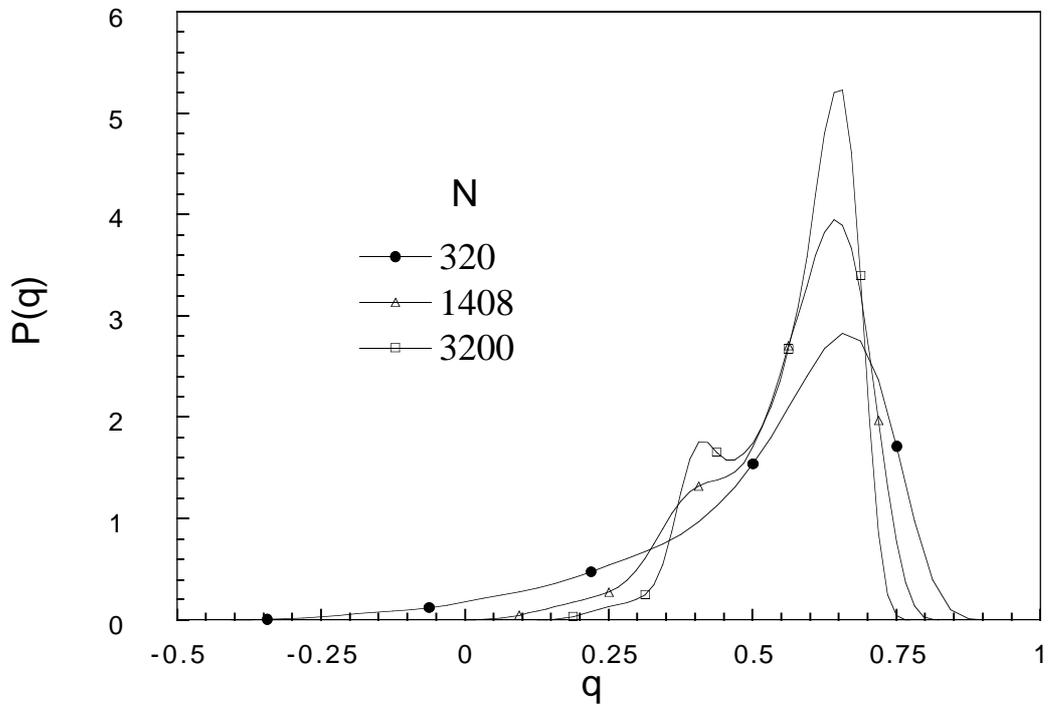
\item[Fig.~1] $P(q)$ for the SK
model at $T=0.5$, $h=0.3$. The error bars are of order 20\% for
$N=3200$ and 15\% for $N=1408$ and less than 5\% for $N=320$. The
symbols are a guide to the eye.
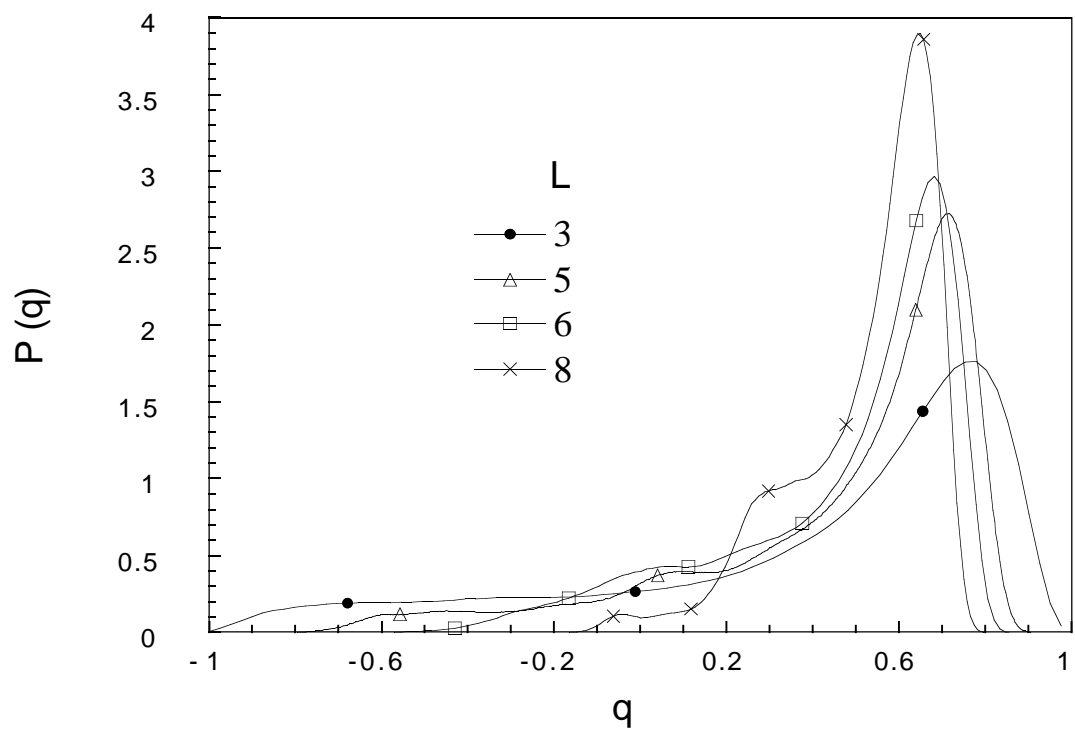
\item[Fig.~2] $P(q)$ for the $4d$ Ising spin glass. Error bars are
smaller than 15\% in all cases. The symbols are a guide to the eye.
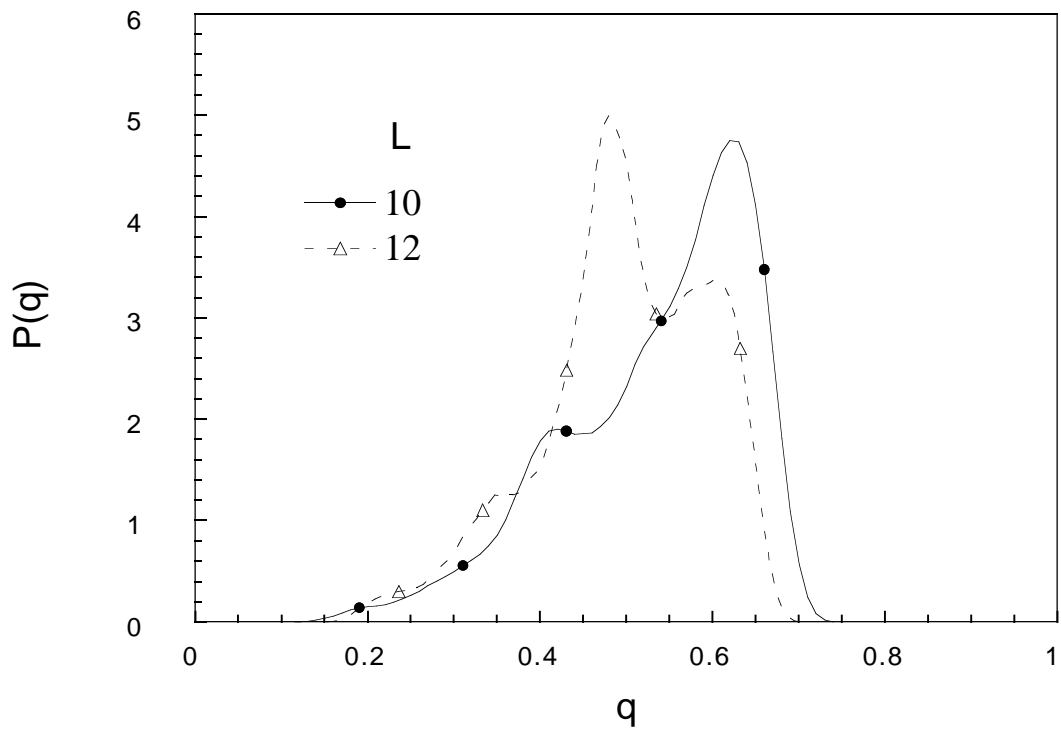
\item[Fig.~3] Non thermalized $P(q)$ for the $4d$ Ising spin glass.
\end{itemize}

\end{document}